\documentclass[UTF8,showpacs,superscriptaddress,reprint, amsmath,amssymb, aps,twocolumn]{revtex4}
\usepackage{amsmath}
\usepackage{graphicx}
\usepackage{dcolumn}
\usepackage{bm}
\usepackage{epsfig}
\usepackage{epsf}
\usepackage{color}
\usepackage{float}
\usepackage[colorlinks=true,
urlcolor=blue,
linkcolor=blue,       
anchorcolor=blue,  
citecolor=blue,        
]{hyperref}

\def \ee {e^+e^-}

\def \half {{1\over 2}}
\def \ee {e^+e^-}

\def \ldc {{\Lambda_c^+}}
\def \ldcb {{\bar\Lambda_c^-}}

\begin{document}

\preprint{APS/123-QED}

\title{Study of parity violation in $\ldc\to\phi p$ and $\ldc \to \omega p$ decays}

\author{Peng-Cheng Hong}
\affiliation{Fudan Uneiversity}
\author{Fang Yan}
\affiliation{Fudan Uneiversity}
\author{Rong-Gang Ping}
\affiliation{Institute of High Energy Physics}
\author{Tao Luo}
\affiliation{Fudan Uneiversity}

\date{\today}

\begin{abstract}
Evidence for CP violation in baryonic decays has not been seen in experiments. With large data events accumulated at $\ee$ collider or Large Hadron Collider, the charmed baryon decays would provide a promising laboratory to test the CP symmetry. In this work, we formulate the $\ldc\to\phi p$ and $ \ldc\to\omega p$ decays for the measurements of their asymmetry parameters in weak decays at the BESIII or the LHCb experiments. The polarization transfer is analyzed in the two processes of $\ee\to\Lambda_c^+\bar\Lambda_c^-$ and $\Lambda_b^0\to \Lambda^+_c\pi^-$, and the formulas of the joint angular distribution for these processes are provided. The sensitivity of the measurement of asymmetry parameters for the above two decay processes is estimated for the future experiments.
\end{abstract}

\maketitle


\section{Introduction}
CP violation plays an important role in the study of weak interactions in the elementary particle physics, and for the attempts to explain the dominance of matter over antimatter in the present universe. First evidence of CP violation was observed in neutral kaon decays \cite{cpkaon}, and then in the $B$ meson decays \cite{cpb1,cpb2}, and recently in the $D$ meson decays \cite{cpd}. But no evidence is found in baryonic decays yet.

Charmed baryonic decays have been used to study weak interaction, and provide a unique laboratory to test CP symmetry. To probe the charmed baryonic decay mechanism, the parity violation was predicted by calculation of decay asymmetry parameters. It has been investigated by a large variety of models, such as pole model, current algebra, SU(3) flavor symmetry and perturbative QCD~\cite{0001,0002,0003,0004,0005,0006,0007,0008,0009}. The decays of charmed baryon, e.g. $\Lambda_c$, receive contributions both from the $W$ emission and $W$ exchange diagrams. Due to the fact that the $W$ emission can be factored out from the short range process, its contribution can be determined from experiments. However, the $W$ exchange contribution, for example in the Cabibbo-favored decay process $\ldc \to\phi  p$ and the singly Cabibbo-suppressed decay process $\ldc \to \omega p$, is quite uncertain in the model predictions. The precise measurement on the asymmetry parameters sheds light on the charmed baryon decay mechanism.

Recently, the branching fraction of $\ldc \to \omega p$ is measured using $pp$ collision data collected at the LHCb experiment \cite{0011} and $\ee$ collision data collected at the Belle detector \cite{0012}, respectively. Using 567 $\mathrm{pb}^{-1}$ data events taken at $\sqrt{s}=4.599$ GeV, the branching fraction of $\ldc \to \phi p$ is measured at the BESIII Collaboration \cite{0010}. However, these statistics are limited to measure the decay asymmetry parameters. The decay rate was intensively investigated by many groups, such as the dynamics calculations \cite{0008,0013,0014,0015} and SU(3) flavour symmetry \cite{,0009,0016,0017,0018,0019,0020}. The branching fraction of $\ldc \to \omega p$ is predicted as $(0.63\pm 0.34)\times 10^{-3}$ from Geng \cite{0019}, $(11.4\pm 5.4)\times 10^{-3}$ from Hsiao \cite{0020} based on SU(3) symmetry,  and $10^{-4}\sim10^{-2}$ from Singer \cite{0015} based on dynamic calculations. A large theoretical uncertainty still exists in the calculation of $W$ exchange diagrams.

It is well known that the parity violation has been observed in the hyperon of hadronic decays, e.g., $\Lambda\to p\pi^-$, in which weak interaction does not conserve the parity. Hence the decay takes place via both the parity allowed $P$ wave and the parity violated $S$ wave for the two-body final states. The interference between the two waves gives rise to the asymmetry angular distribution, which is characterized with an intrinsic asymmetry parameter, for example, $\alpha_\Lambda = 0.732\pm 0.004$ \cite{alpha_lam} for $\Lambda\to p\pi^-$.

Similarly, the asymmetry parameters of $\Lambda_c\to Vp~(V=\phi,\omega)$ decays encode partly the weak interactions in the $\Lambda_c$  decays. We define the parameters with helicity amplitudes $B_{\lambda_3,\lambda_4}$, here $\lambda_3$ and $\lambda_4$ are the helicity values of vector meson and proton, respectively. The nonzero value of helicity amplitude assumes four configurations, namely, $B_{1,\half},~B_{-1,-\half},~B_{0,\half}$ and $B_{0,-\half}$. If parity conserves in the decay, the amplitudes of four components are reduced to two independent amplitudes with relation $B_{1,\half}=B_{-1,-\half},~B_{0,\half}=B_{0,-\half}$. In other words, if $B_{1,\half}\neq B_{-1,-\half}$ or/and $B_{0,\half}\neq B_{0,-\half}$, it indicates the parity violation takes place in the decay. We introduce three parameters to identify the possible effects due to the parity violation in the $\Lambda_c\to V p$ decays, i.e.
\begin{eqnarray}\label{def.1}
	\alpha_{\phi p}&=&\frac{|B_{1,\half}|^2-|B_{-1,-\half}|^2} {|B_{1,\half}|^2+|B_{-1,-\half}|^2},\nonumber\\
	\beta_{\phi p}&=&\frac{|B_{0,\half}|^2-|B_{0,-\half}|^2} {|B_{0,\half}|^2+|B_{0,-\half}|^2},\nonumber \\
	\gamma_{\phi p}&=&\frac{|B_{1,\half}|^2+|B_{-1,-\half}|^2}{|B_{0,\half}|^2+|B_{0,-\half}|^2}.
\end{eqnarray}

Obviously, the nonzero values for $\alpha_{\phi p}$ and $\beta_{\phi p}$ imply the P parity violation in this decay, which is more straightforward than the definition of asymmerty parameters with two partial wave amplitudes introduced by Lee-Yang \cite{0021}.
The two sets of parameter definitions can certainly convert to each other via the relation between the helicity decay amplitude and the partial-wave amplitude with Clebsch-Gordan coefficient.

Experimentally, there is another definition for parametrization of the parity asymmetry analogous to the case of  $\Lambda_b \to \phi p$ \cite{0022}, i.e.
\begin{equation}\label{Eq2}
	\Lambda_{\phi p}={|B_{1,\half}|^2-|B_{-1,-\half}|^2+|B_{0,\half}|^2-|B_{0,-\half}|^2\over |B_{1,\half}|^2+|B_{-1,-\half}|^2+|B_{0,\half}|^2+|B_{0,-\half}|^2}.
\end{equation}
This asymmetry parameter is insensitive to test the asymmetry effects due to the longitudinal or transversal polarition of the vector mesons $\phi$ or $\omega$.
For example,  if $\alpha_{\phi p}$ and $\beta_{\phi p}$  are nonzero, but they have reverse sign and the same absolute values, then the former definition can identify the asymmetry properties, while the latter one can not, which has zero value as the case of parity conservation.
In this paper, we use Eq.~(\ref{def.1}) to identify the parity violation effects in the decay $\Lambda_c \to pV (V=\omega, \phi)$ .
If we choose the amplitude reference as $|B_{1,\half}|^2=1$, then it relates to the parametrization in Eq.~(\ref{Eq2}) as
\begin{equation}
	\Lambda_{\phi p}={\beta_{\phi p}+\alpha_{\phi p}\gamma_{\phi p}\over 1+\gamma_{\phi p}},
\end{equation}
and we reformulate the amplitudes in terms of magnitude and phase angles as $B_{\lambda_{3},\lambda_4}=b_{\lambda_{3},\lambda_4}e^{i\xi_{\lambda_{3},\lambda_4}}$, then we have
\begin{eqnarray}\label{wkamp}
	b_{1,\half}^2&=&1,~~~b_{-1,-\half}^2={1-\alpha_{\phi p}\over 1+\alpha_{\phi p}},\nonumber\\
	b_{0,\half}^2&=&{1+\beta_{\phi p}\over (1+\alpha_{\phi p})\gamma_{\phi p}},~~~b_{0,-\half}^2={1-\beta_{\phi p}\over (1+\alpha_{\phi p})\gamma_{\phi p}}.
\end{eqnarray}

Assume that the CP symmetry is a good approximation in the $\ldc$ and $\ldcb$ decays, then we have the asymmetry parameters for the conjugate $\ldcb$ decay as
\begin{equation}
	\alpha_{\phi p}=-\overline{\alpha_{\phi p}},~~~\beta_{\phi p}=-\overline{\beta_{\phi p}},~~~\gamma_{\phi p}=\overline{\gamma_{\phi p}}.
\end{equation}
Thus, we only formulate the $\Lambda_c^+$ decays as follows.
Experimentally, extraction of these parameters can be performed by fitting the data events with the join angular distributions as done in Refs.~\cite{0023,0024}.
Once more data is available, the two conjugate  decays can studied separately for further test of the CP symmetry in charmed baryonic decays.
\section{$\ldc\to V p$}
\subsection{Kinematic variable and helicity amplitude}
Let's first consider the decay process $\ldc\to V p$, where V denotes $\phi$ or $\omega$, and the subsequent decays $\phi\to K^+K^-$ and $\omega\to\pi^+\pi^-\pi^0$.
We formulate the decays with helicity formalism, from which the asymmetry decay parameters are introduced and can be extracted from fitting the data events with the functions of angular distributions.

\begin{figure}[H]
\includegraphics[width=0.45\textwidth]{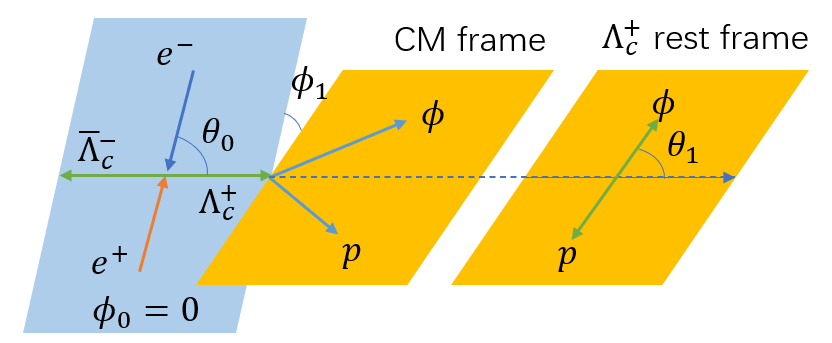}
\centering
	\caption{Definition of helicity frame at the BESIII experiment.}
	\label{fig:angle_BESIII}
\end{figure}

Helicity angles for $\ldc\to \phi p, \phi \to K^+K^-$ are defined as shown in Fig.~\ref{fig:angle_BESIII} and the corresponding helicity amplitutes are tabulated in Table~\ref{def.2}.
For strong decay which meets the parity conservation, one has
\begin{eqnarray}
	F^J_{\lambda_m, \lambda_n}=\eta\eta_1\eta_2(-1)^{J-s_1-s_2} F^J_{-\lambda_m, -\lambda_n},
\end{eqnarray}
where $J$ is the spin of the mother particle, $s_i$ is the spin of daughter particles, $\eta_i$ is the intrinsic parity of each particle and $\lambda_i$ indicates the helicity values \cite{0025}.
For this decay, we have
$J=1,s_1=s_2=0$, $\eta=-1,\eta_1=\eta_2=-1$, and the helicity of Kaon is zero, with only one independent amplitude exists, which is labeled as $C$.
\begin{table}[htbp]
	\caption{Definition of $\phi \to K^+ K^-$ decays, helicity angles and amplitudes, where $\lambda_i$ indicates the helicity values for the corresponding hadron.}\label{def.2}
	\begin{tabular}{lll}
		\hline\hline
		decay & helicity angle & helicity amplitude \\
		\hline
		$\ldc(\lambda_1)\to\phi(\lambda_3)p(\lambda_4)$ & ($\theta_1,\phi_1$)& $B_{\lambda_3,\lambda_4}$\\
		\hline
		$\phi(\lambda_3)\to K^+K^-$ &($\theta_2,\phi_2$)&$C$\\
		\hline\hline
	\end{tabular}
\end{table}

\begin{table}[htbp]
	\caption{Definition of $\omega \to \pi^+ \pi^- \pi^0$ decays, helicity angles and amplitudes.}\label{def.3}
	\begin{tabular}{lll}
		\hline\hline
		decay & helicity angle & helicity amplitude \\
		\hline
		$\ldc(\lambda_1)\to\omega(\lambda_3)p(\lambda_4)$ & ($\theta_1,\phi_1$)& $B_{\lambda_3,\lambda_4}$\\
		\hline
		$\omega(\lambda_3)\to \pi^+\pi^-\pi^0$ &($\alpha,\beta,\gamma$)&$C$\\
		\hline\hline
	\end{tabular}
\end{table}

For the $\omega \to \pi^+\pi^-\pi^0$ decay, helicity angles and amplitudes are defined in the same way as shown in Table~\ref{def.3}.
If P parity is conseverd in three-body systems, we have a symmetry relation for helicity amplitudes as
\begin{eqnarray}
	C^J_\mu(E_i \lambda_i)=\eta\eta_1\eta_2\eta_3(-1)^{s_1+s_2+s_3+\mu}C^J_\mu(E_i -\lambda_i),
\end{eqnarray}
where $J$ is the spin of the mother particle, $s_1,s_2,s_3$ are for the final particles, $\mu$ is the eigenvalue of $J$ in the body-fixed z-axis, $E_i$ and $\lambda_i$ are the energy and helicity of the three particles \cite{0025}. For pion, its helicity is zero,  therefore we can have
\begin{eqnarray}
	C^1_1(E_i)=-C^1_1(E_i)=0,~~~C^1_{-1}(E_i)=-C^1_{-1}(E_i)=0,
\end{eqnarray}
with only one helicity amplitude $C^1_0(E_i)$ exists, and we denote it as $C$ as well, whose form has no difference compared with  $\phi$ in joint helicity deacy amplitudes.

\subsection{Spin density matrix}
The spin density matrix (SDM) encodes the whole polarization information of a particle in a decay, and can be used to demonstrate the spin and polarization transfer in the sequential decays clearly \cite{0031}.
The SDM of spin-$1\over2$ particles, like $\ldc$, can be expressed as
\begin{eqnarray}
	\rho={{\bm{\bm{\mathcal{P}_0}}}\over2}(\bm{I}+\bm{\mathcal{P}}\cdot\bm{\sigma}),
\end{eqnarray}
where $\bm{\mathcal{P}_0}$ is the unpolarized cross section, $\bm{\mathcal{P}}$ is the polarized vector, and $\bm{I}$ is a 2$\times$2 unit matrix, and $\bm{\sigma}$ is the Pauli matrix as we all know.
Specifically, we can write the SDM of $\ldc$ as
\begin{eqnarray}\label{pola define}
	\rho^\ldc={{\bm{\mathcal{P}_0}}\over2}{\left(\begin{array}{cc}1+\bm{\mathcal{P}_z}&\bm{\mathcal{P}_x}-i\bm{\mathcal{P}_y}  \\\bm{\mathcal{P}_x}+i\bm{\mathcal{P}_y}& 	1-\bm{\mathcal{P}_z}\end{array}\right)},
\end{eqnarray}
where $\bm{\mathcal{P}_x}$ and $\bm{\mathcal{P}_y}$ is the transverse polarization and $\bm{\mathcal{P}_z}$ is the longitudinal polarization in the coordinate system as defined in Fig.~\ref{fig:angle_BESIII}.

Using the SDM of $\ldc$, the elements of $\phi$ or $\omega $ SDM can be calculated as
\begin{eqnarray}
	\rho^\phi_{\lambda_3,\lambda_3}\propto\sum_{\lambda_{1},\lambda_{1}^{'},\lambda_4}{\rho^{\ldc}_{\lambda_{1},\lambda_{1}^{'}}}D^{{1\over2}*}_{\lambda_{1},\lambda_{3}-\lambda_{4}}(\phi_1,\theta_1,0)\nonumber\\
	D^{{1}}_{\lambda_{1}^{'},\lambda_{3}^{'}-\lambda_{4}}(\phi_1,\theta_1,0)B_{\lambda_{3},\lambda_{4}}B_{\lambda_{3}^{'},\lambda_{4}}^{*}.
\end{eqnarray}
where $D^{J}_{\lambda_i,\lambda_k}$ is the Wigner-D function. The result of this calculation is listed in Appendix.~\ref{SDM of phi}.

On the other hand, the SDM can be rewritten in terms of the real multipole parameter, $r^L_M$, which is defined as
\begin{equation}
	r^L_M=\mathrm{Tr}[\rho^\phi \cdot Q^L_M],
\end{equation}
where $Q^L_M$ is a series of Hermitian basis matrices and can be calculated by the method in Ref.\cite{0026}.
Tr denotes taking a trace of the matrix.
The $L$-rank index runs from 1 to 2$J$, and $M$ is regarded as an integer number runs from -$L$ to $L$. For a spin-$J$ particle, the SDM can be expressed as
\begin{eqnarray}
	\rho={{r^0_0}\over{2J+1}}(\bm{I}+2J\sum\limits_{L=1}^{2J}\sum\limits_{M=-L}^{L}r^L_{M}Q^L_{M}),
\end{eqnarray}
where $\bm{I}$ denotes a unit matrix with dimension of $2J+1$ and $r^0_0$ corresponds to the unpolarized decay rate with $r^0_0 = \mathrm{Tr}\rho$, then we get the SDM of $\phi$ as
\begin{widetext}
	\begin{eqnarray}
		\rho^\phi=\frac{{r}^0_0}{3}\left(
		\begin{array}{ccc}
			1+\sqrt{3} {r}^1_0+{r}^2_0
			&\sqrt{\frac{3}{2}}(-i {r}^1_{-1}
			+ {r}^1_1
			-i{r}^2_{-1}+ {r}^2_1)
			& \sqrt{3}(-i{r}^2_{-2}+{r}^2_2) \\
			\sqrt{\frac{3}{2}}(	i{r}^1_{-1}
			+ {r}^1_1
			+i{r}^2_{-1}
			+{r}^2_1 )
			&1-2 {r}^2_0
			&\sqrt{\frac{3}{2}}(-i {r}^1_{-1}
			+{r}^1_1
			+i {r}^2_{-1}-{r}^2_1)\\
			\sqrt{3}(i{r}^2_{-2}
			+ {r}^2_2)
			&\sqrt{\frac{3}{2}}(i {r}^1_{-1}
			+{r}^1_1
			-i{r}^2_{-1}
			-{r}^2_1 )
			&1-\sqrt{3} {r}^1_0
			+{r}^2_0
		\end{array}
		\right).
	\end{eqnarray}
\end{widetext}
The multipole parameters can be related to the polarization expressions in Appendix.~\ref{mul_para},

In the equations of these appendices, we have reformulated the helicity amplitude for $\ldc\to p\phi$ in terms of magnitude and phase angles, namely,
\begin{eqnarray}
	B_{\lambda_3,\lambda_4}&=&b_{\lambda_3,\lambda_4}e^{i\xi_{\lambda_3,\lambda_4}},~~~\Delta_{01}=\xi _{0,\frac{1}{2}}-\xi_{1,\frac{1}{2}},\nonumber\\
	\Delta_{10}&=&\xi _{-1,-\frac{1}{2}}-\xi _{0,-\frac{1}{2}},
\end{eqnarray}
and the modulus $b_{\lambda_{3},\lambda_4}$ relates to the asymmetry parameters according to Eq.~(\ref{wkamp}).


\subsection{Joint angular distribution}
By making use of the individual helicity amplitude of the decay in a chain, namely, $\ldc\to\phi p$ and $\phi\to K^+K^-$, the joint angular distribution of this process is expanded in terms of multipole parameters $r^L_M$.
The square of amplitude $|C|^2$ can be normalized to one, it reads
\begin{eqnarray}\label{Joint_ang_dis}	
	\mathcal{W}(\theta_0,\theta_1,\theta_2,\phi_1,\phi_2)&\propto&\sum_{\lambda_3,\lambda_3^{'}}\rho^\phi_{\lambda_3,\lambda_3^{'}}D^{1*}_{\lambda_3,0}D^{1}_{\lambda_3^{'},0}|C|^2\nonumber\\
	&=&-\frac{1}{6} r^0_0 \left\{2 \sqrt{3} \mathrm{\sin}2\theta_2[r^2_{-1} \mathrm{\sin} \phi_2\right.\nonumber\\
	&+&r^2_1 \mathrm{\cos} \phi_2]+3 r^2_0 \mathrm{\cos}2\theta_2\nonumber\\
	&+&r^2_0-2\left.\right\}.
\end{eqnarray}

The helicity amplitude prescription of the $\omega \to \ \pi^+\pi^-\pi^0$ process is different from the $\phi \to K^+ K^-$ process.
In the $\omega \to \pi^+\pi^-\pi^0$ decay, the Euler angles, $(\alpha, \beta, \gamma)$ is used, which describe the rotation to carry the $\omega$ rest frame to overlap the helicity frame formed by the three pion momenta.
The deduction of angular distribution is similar as the $\phi$ decay.
The result is
\begin{eqnarray}
	\mathcal{W}(\theta_0,\theta_1,\phi_1,\alpha,\beta,\gamma)&\propto& -\frac{1}{6} r^0_0 \left\{2 \sqrt{3} \mathrm{\sin}2\beta [r^2_{-1} \mathrm{\sin} \alpha\right.\nonumber\\
	&+&r^2_1 \mathrm{\cos} \alpha]+3 r^2_0 \mathrm{\cos}2\beta\nonumber\\
	&+&r^2_0-2\left.\right\}.
\end{eqnarray}

\subsection{Polarization observable}
The helicity angles in the $\phi \to K^+ K^-$ process can be used to form a polarization observable to describe the relevant multipole parameters \cite{0023}.
We give the first moments of Wigner D-function as follows to display the $\phi$ polarization effects(the $\omega$ decay is nearly the same).
\begin{eqnarray}
	\langle \mathrm{\cos}2\theta_2\rangle &=&-{1\over 15}(5+8r^2_0),\\
	\langle \mathrm{\sin}2\theta_2\mathrm{\cos}\phi_2 \rangle &=&-{4\over 5\sqrt 3}r^2_1,\\
	\langle \mathrm{\sin}2\theta_2\mathrm{\sin}\phi_2 \rangle &=&-{4\over 5\sqrt 3}r^2_{-1}.
\end{eqnarray}

\section{$\ldc$ production}
Let's consider the cascade decay, i.e. $\ee\to\gamma^*(M_0)\to\ldc(\lambda_{1})\ldcb(\lambda_{2}),
\ldc(\lambda'_{1})\to\phi(\lambda_3)p(\lambda_4)$,
$\phi(\lambda_3)\to K^+K^-$, which can be studied at the BESIII experiment, and $pp\to\Lambda_b^0X,~ {{\Lambda}_b^0}(M_0)\to\ldc(\lambda_1)\pi^-,\ldc(\lambda_1^{'})\to\phi(\lambda_3) p(\lambda_4)$, $\phi(\lambda_3^{'})\to K^+K^-$, which can be studied at the LHCb experiment, where the helicity values, i.e. $M_0,\lambda_i$ and $\lambda'_j$, are indicated in the parenthesis for the corresponding particles. Table \ref{def.4} tabulates the definition of helicity angels, amplitudes in producing $\ldc$ at the BESIII experiment and the LHCb experiment.
\begin{table}[htbp]
	\caption{Definition of helicity angles and amplitudes in experiments.}\label{def.4}
	\begin{tabular}{lll}
		\hline\hline
		decay & helicity angle & helicity amplitude \\\hline
		$\gamma^*(M_0)\to\ldc(\lambda_1)\ldcb(\lambda_2)$ & ($\theta_0,\phi_0$) & $A_{\lambda_1,\lambda_2}$\\\hline
		$\Lambda_b^0(M_0)\to\ldc(\lambda_1)\pi^-$ &($\theta_0,\phi_0$)&$A_{\lambda_1}$ \\
		\hline\hline
	\end{tabular}
\end{table}

\subsection{$\ee annihilation$}
At the BESIII experiment, the baryon pairs are produced from the unpolarized beam $\ee$ collisions \cite{0030}, the decay chain is $\ee\to\ldc\ldcb, \ldc\to\phi p$, $\phi\to K^+K^-$ and as shown in Fig.~\ref{fig:angle_BESIII}.
In this decay, the polar and azimuthal angles are defined as follows
\begin{eqnarray}
	\phi_0&=&0,~~~~~~~~~~~\mathrm{\cos}\theta_0=\frac{\vec{p}_{e^+}\cdot{\vec{p}_\ldc}}{\lvert\vec{p}_{e^+}\lvert\cdot\lvert\vec{p}_\ldc\rvert},\nonumber\\
	\mathrm{\cos}\phi_1&=&|\vec{n}_1\cdot{\vec{n}_2}|,	~~~\mathrm{\cos}\theta_1=\frac{\vec{p}_\phi\cdot{\vec{p}_\ldc}}{\lvert\vec{p}_\phi\lvert\cdot\lvert\vec{p}_\ldc\rvert},\nonumber\\
	\mathrm{\cos}\phi_2&=&|\vec{n}_2\cdot{\vec{n}_3}|,
	~~~\mathrm{\cos}\theta_2=\frac{\vec{p}_{K^+}\cdot{\vec{p}_\phi}}{\lvert\vec{p}_{K^+}\lvert\cdot\lvert\vec{p}_\phi\rvert},
\end{eqnarray}
where $\vec{p}$ denotes the momentum of each particle in the rest frame of their respective parent particles, and $\vec{n}_i$ is the normal vector defined as
\begin{eqnarray}
	\vec{n}_1&=&\frac{\vec{p}_{e^+}\times{\vec{p}_\ldc}}{\lvert\vec{p}_{e^+}\lvert\cdot\lvert\vec{p}_\ldc\rvert\cdot{\mathrm{\sin}\theta_0}},
	~~~\vec{n}_2=\frac{\vec{p}_{\ldc}\times{\vec{p}_\phi}}{\lvert\vec{p}_{\ldc}\lvert\cdot\lvert\vec{p}_\phi\rvert\cdot{\mathrm{\sin}\theta_1}},\nonumber\\
	\vec{n}_3&=&\frac{\vec{p}_{\phi}\times{\vec{p}_{K^+}}}{\lvert\vec{p}_{\phi}\lvert\cdot\lvert\vec{p}_{K^+}\rvert\cdot{\mathrm{\sin}\theta_2}}.
\end{eqnarray}
For example, $\theta_1$ is the polar angle between the momentum of $\phi$ and $\ldc$ in $\ldc$ rest frame, and $\phi_2$ is the azimuthal angle between the $\ldc\to\phi p$ decay plane and the $\phi\to K^+K^-$ decay plane as shown in Fig.~\ref{fig:angle_BESIII}.

The SDM of $\ldc$ for BESIII experiment can be formulated as
\begin{eqnarray}
	\rho^\ldc_{\lambda_1,\lambda_1^{'}}&\propto&\sum_{{M_0},\lambda_2}{\rho^{\gamma^*}_{M_0,M_0^{'}}}D^{1*}_{M_0,\lambda_{1}-\lambda_{2}}(\phi_0,\theta_0,0)\nonumber\\
	&\times&D^{{1}}_{M_0^{'},\lambda_{1}^{'}-\lambda_{2}}(\phi_0,\theta_0,0)A_{\lambda_{1},\lambda_{2}}A_{\lambda_{1}^{'},\lambda_{2}}^{*},
\end{eqnarray}
where ${\rho^{\gamma^*}_{M_0,M_0^{'}}}$ is a diagonal matrix written as $\frac{1}{2}\mathrm{diag}\left\{\right.1,0,1\left.\right\}$.
For the parity conserved decay, one has the symmetry relation as
\begin{eqnarray}\label{symmetry Amp}
	A_{-\half,-\half}=A_{\half,\half},~A_{-\half,\half}=A_{\half,-\half}.
\end{eqnarray}

It is determined that the unpolarized cross section is $\bm{\mathcal{P}_0}=1+\alpha_c\mathrm{\cos}^2\theta_0$,
where $\alpha_c$ is the angular distribution parameter for $\ee\to\ldc\ldcb$, and is defined as
\begin{eqnarray}
	\alpha_c={|A_{\half,-\half}|^2-2|A_{\half,\half}|^2\over |A_{-\half,\half}|^2+2|A_{\half,\half}|^2}.
\end{eqnarray}
The polarization of $\ldc$ is calculated to be
\begin{eqnarray}
	\bm{\mathcal{P}_y}={\sqrt{1-\alpha_c^2}\mathrm{\sin}(2\theta_0)\mathrm{\sin}\Delta \over 2(1+\alpha_c\mathrm{\cos}^2\theta_0)},	~\bm{\mathcal{P}_x}=	\bm{\mathcal{P}_z}=0,
\end{eqnarray}
where $\Delta$ is a parameter to describe the phase angle difference between the two independent amplitudes for this process, and we leave out the constant $|A_{-\half,\half}|^2+2|A_{\half,\half}|^2$ in $\bm{\mathcal{P}_0}$ expression.
Consequently, $\ldc$ may have a transverse polarization spontaneously  from unpolarized $\ee$ if $\Delta$ is nonzero.

\subsection{$\Lambda_b^0 \to \ldc \pi^-$}
\begin{figure}[H]
	\includegraphics[width=0.45\textwidth]{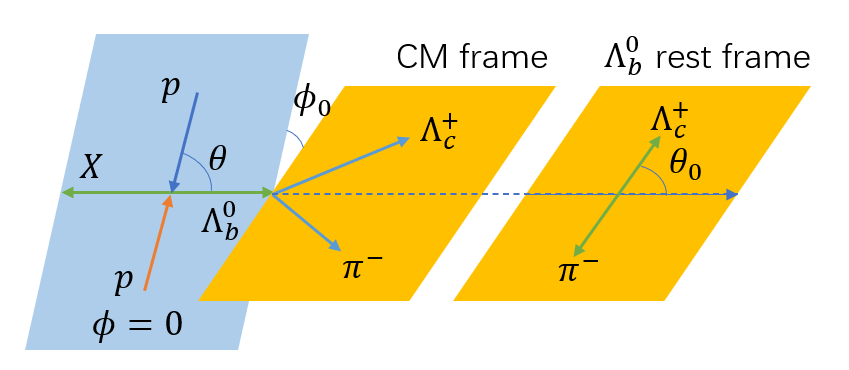}
	\centering
	\caption{Definition of helicity frame at the LHCb experiment.}
	\label{fig:angle_LHCb}
\end{figure}

At the LHCb experiment, assuming the $\ldc$ is produced by the polarized ${\Lambda}_b^0$ \cite{0028,0029}, we write the decay chain as ${\Lambda}_b^0\to\ldc\pi^-, \ldc\to\phi p$, $\phi\to K^+K^-$.
In this decay chain, we need to define two extra angles and a normal vector as
\begin{eqnarray}
	\mathrm{\cos}\phi_0&=&\frac{\vec{n}_0\cdot{\vec{n}_1}}{\lvert\vec{n}_0\lvert\cdot\lvert\vec{n}_1\rvert},
	~~~\vec{n}_0=\frac{\vec{p}_p\cdot{\vec{p}_{{\Lambda}_b^0}}}{\lvert\vec{p}_p\rvert\cdot{\lvert{\vec{p}_{{\Lambda}_b^0}}\rvert}\mathrm{\sin}\theta},\nonumber\\
	\mathrm{\cos}\theta&=&\frac{\vec{p}_{p}\cdot{\vec{p}_{{{\Lambda}}_b^0}}}{\lvert\vec{p}_{p}\lvert\cdot\lvert\vec{p}_{{{\Lambda}}_b^0}\rvert},
\end{eqnarray}
here the angles are shown in Fig.~\ref{fig:angle_LHCb}, compared to the $\ldc$ produced in $\ee$ collision.
The definitions of other angles are same as those at the BESIII experiment.
Consequently, these quantities can be constructed with the momentum registered in detectors.

According to Eq.~(\ref{pola define}), the SDM of ${\Lambda}_b^0$ can be written as
\begin{eqnarray}
	\rho^{{\Lambda}_b^0}={{\bm{\mathcal{P}_0}^b}\over2}{\left(\begin{array}{cc}1+\bm{\mathcal{P}_z}^b&\bm{\mathcal{P}_x}^b-i\bm{\mathcal{P}_y}^b  \\\bm{\mathcal{P}_x}^b+i\bm{\mathcal{P}_y}^b& 1-\bm{\mathcal{P}_z}^b\end{array}\right)}.
\end{eqnarray}
The SDM of decayed particle $\ldc$ can be formulated as
\begin{eqnarray}
	\rho^\ldc_{\lambda_1,\lambda_1^{'}}&\propto&\sum_{{M_0},\lambda_2}{\rho^{{\Lambda_b^0}}_{M_0,M_0^{'}}}D^{\frac{1}{2}*}_{M_0,\lambda_{1}-\lambda_{2}}(\phi_0,\theta_0,0)\nonumber\\
	&\times&D^{\frac{1}{2}}_{M_0^{'},\lambda_{1}^{'}-\lambda_{2}}(\phi_0,\theta_0,0)A_{\lambda_{1}}A_{\lambda_{1}^{'}}^{*}.
\end{eqnarray}
We define an asymmetry parameter to express the degree of P violation in this weak decay as
\begin{eqnarray}
	\alpha_b={|A_{\half}|^2-|A_{-\half}|^2\over |A_{\half}|^2+|A_{-\half}|^2},
\end{eqnarray}
The unpolarized cross section and the polarization of $\ldc$ can be expressed as
\begin{eqnarray}
	\bm{\mathcal{P}_0}&=&\frac{\bm{\mathcal{P}_0}^b}{2}(1+\alpha_b{\bm{\mathcal{P}_z}^b}\mathrm{\cos}\theta_0+	\alpha_b{\bm{\mathcal{P}_x}^b}\mathrm{\cos}\phi_0\mathrm{\sin}\theta_0\nonumber\\
	&+&	\alpha_b{\bm{\mathcal{P}_y}^b}\mathrm{\sin}\theta_0\mathrm{\sin}\phi_0),\nonumber\\
	\bm{\mathcal{P}_0}\bm{\mathcal{P}_x}&=&\frac{\sqrt{1-\alpha_b^2}}{2}\bm{\mathcal{P}_0}^b(-\bm{\mathcal{P}_y}^b\mathrm{\cos}\phi_0\mathrm{\sin}\Delta_b\nonumber\\
	&+&\bm{\mathcal{P}_x}^b\mathrm{\cos}\theta_0\mathrm{\cos}\phi_0\mathrm{\cos}\Delta_b-\bm{\mathcal{P}_z}^b\mathrm{\sin}\theta_0\mathrm{\cos}\Delta_b\nonumber\\
	&+&\bm{\mathcal{P}_x}^b\mathrm{\sin}\phi_0\mathrm{\sin}\Delta_b+\bm{\mathcal{P}_y}^b\mathrm{\cos}\theta_0\mathrm{\sin}\phi_0\mathrm{\cos}\Delta_b),\nonumber\\
	\bm{\mathcal{P}_0}\bm{\mathcal{P}_y}&=&\frac{\sqrt{1-\alpha_b^2}}{2}\bm{\mathcal{P}_0}^b(\bm{\mathcal{P}_y}^b\mathrm{\cos}\phi_0\mathrm{\cos}\Delta_b\nonumber\\
	&+&\bm{\mathcal{P}_x}^b\mathrm{\cos}\theta_0\mathrm{\cos}\phi_0\mathrm{\sin}\Delta_b-\bm{\mathcal{P}_z}^b\mathrm{\sin}\theta_0\mathrm{\sin}\Delta_b\nonumber\\
	&-&\bm{\mathcal{P}_x}^b\mathrm{\sin}\phi_0\mathrm{\cos}\Delta_b+\bm{\mathcal{P}_y}^b\mathrm{\cos}\theta_0\mathrm{\sin}\phi_0\mathrm{\sin}\Delta_b),\nonumber\\
	\bm{\mathcal{P}_0}\bm{\mathcal{P}_z}&=&\frac{\bm{\mathcal{P}_0}^b}{2}(\alpha_b+{\bm{\mathcal{P}_z}^b}\mathrm{\cos}\theta_0+{\bm{\mathcal{P}_x}^b}\mathrm{\cos}\phi_0\mathrm{\sin}\theta_0\nonumber\\
	&+&{\bm{\mathcal{P}_y}^b}\mathrm{\sin}\theta_0\mathrm{\sin}\phi_0).
\end{eqnarray}

In these equations, $\Delta_b$ is a parameter describing the phase angle difference between the two independent amplitudes for this process, and we have ignored the constant $|A_{\half}|^2+|A_{-\half}|^2$ on the right side of the four equations above. These formulas show the transfer process of  polarization in the decay chain.
To obtain the polarization effects, we just need to replace the elements of the multipole parameters in Appendix.~\ref{mul_para}.

\section{sensitivity of asymmetry parameters measurements}
Sensitivity estimation is indispensable when data taken proposal is suggested or physics program is designed before the experiment is carried out. We estimate the statistical sensitivity in terms of the number of observed events for measurement of the $\Lambda_c$ decay asymmetry parameter. To increase the sensitivity, the information of the full spin transfer through the complete decay chain is used, rather than only the information from the $\Lambda_c$ decay. 
Comparing the joint angular distribution of these two decays of $\Lambda_c \to \phi p$ and $\Lambda_c \to \omega p$, one can see that they takes the same form by making the replacement of $(\theta_2,\phi_2)\to (\beta,\alpha)$ in the $\phi p$ channel. Therefore, we only consider the $\omega p$ channel in our estimation with an assumption of different asymmetry parameters, which can be extracted from fitting the normalized angular distribution with adequate data samples. The angular distribution $\widetilde{\mathcal{W}}(\theta_0,\theta_1,\phi_1,\alpha,\beta,\gamma)$ is defined as
\begin{eqnarray}
	\widetilde{\mathcal{W}}=\frac{\mathcal{W}(\theta_0,\theta_1,\phi_1,\alpha,\beta,\gamma)}{\int\cdot\cdot\cdot\int\mathcal{W}(\cdot\cdot\cdot)\mathrm{d}\mathrm{\cos}\theta_0\mathrm{d}\mathrm{\cos}\theta_1\mathrm{d}\mathrm{\cos}\beta \mathrm{d}\phi_1d\alpha \mathrm{d}\gamma}.
\end{eqnarray}
Here the denominator plays a role of normalization. For a set of data samples, a likelihood function is defined as
\begin{eqnarray}
	L=\prod_{i=1}^{N}\widetilde{\mathcal{W}}(\theta_0,\theta_1,\phi_1,\alpha,\beta,\gamma),
\end{eqnarray}
where $N$ is the number of observed events \cite{0027}.
Based on the maximum likelihood method, the estimated statistical sensitivity corresponding to each decay parameter  is determined by the relative uncertainty
\begin{eqnarray}\label{delta_alpha}
	\delta(\alpha_{\omega p})=\frac{\sqrt{V(\alpha_{\omega p})}}{|\alpha_{\omega p}|},
\end{eqnarray}
where $V(\alpha_{\omega p})$ is the variance of the decay parameter $\alpha_{\omega p}$, which can be calculated by
\begin{eqnarray}\label{V_alpha}
	V^{-1}(\alpha_{\omega p})&=&N\int\frac{1}{\widetilde{\mathcal{W}}}[\frac{\partial	\widetilde{\mathcal{W}}}{\partial\alpha_{\omega p}}]^2\mathrm{d}\mathrm{\cos}\theta_0\mathrm{d}\mathrm{\cos}\theta_1\mathrm{d}\mathrm{\cos}\beta\nonumber\\
	&\times&\mathrm{d}\phi_1\mathrm{d}\alpha \mathrm{d}\gamma.
\end{eqnarray}
The other two parameters $\delta_{\beta_{\omega p}}$ and $\delta_{\gamma_{\omega p}}$ are defined analogously to the Eq.~({\ref{delta_alpha}}) and Eq.~(\ref{V_alpha}).\\

\begin{figure}[H]
	\includegraphics[width=0.5\textwidth]{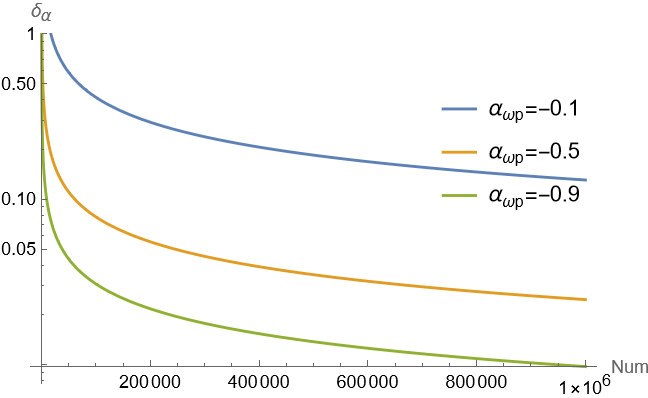}
	\centering
	\caption{The $\alpha_{\omega p}$ sensitivities relative to signal yields N in terms of different value $\alpha_{\omega p}$.}
	\label{fig:alpha}
\end{figure}

To figure out the dependence of sensitivity on the signal yields $N$ produced at the $\ee$ collider experiment, we compute the value of $\delta(\alpha_{\omega p})$ by naively taking the parameters as
\begin{eqnarray}
	\alpha_{c}&=&0.2,~~~\beta_{\omega p}=0.1,~~~\gamma_{\omega p}=0.5,\nonumber\\
	\Delta&=&\frac{\pi}{3},~~~~\Delta_{01}=\frac{\pi}{4},~~~~\Delta_{10}=\frac{\pi}{6}.
\end{eqnarray}
We plot the sensitivities of a set of different $\alpha_{\omega p}$ values versus the statistics, as shown in Fig.~\ref{fig:alpha}  , where the correlative experimental effects, such as the backgrounds level and the detection efficiencies are not considered.

\begin{figure}[H]
	\includegraphics[width=0.5\textwidth]{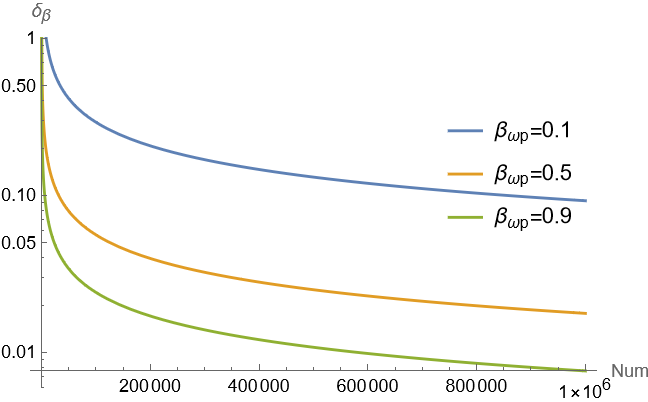}
	\centering
	\caption{The $\beta_{\omega p}$ sensitivities relative to signal yields N in terms of different value $\beta_{\omega p}$.}
	\label{fig:beta}
\end{figure}

\begin{figure}[H]
	\includegraphics[width=0.45\textwidth]{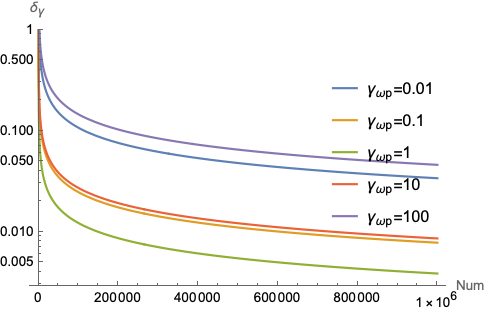}
	\centering
	\caption{The $\gamma_{\omega p}$ sensitivities relative to signal yields N in terms of different value $\gamma_{\omega p}$.}
	\label{fig:gamma}
\end{figure}

By taking the parameters as
\begin{eqnarray}
	\alpha_{c}&=&0.2,~~~\alpha_{\omega p}=-0.5,~~~\gamma_{\omega p}=0.5,\nonumber\\
	\Delta&=&\frac{\pi}{3},~~~~\Delta_{01}=\frac{\pi}{4},~~~~~~\Delta_{10}=\frac{\pi}{6},
\end{eqnarray}
and
\begin{eqnarray}
	\alpha_{c}&=&0.2,~~~\alpha_{\omega p}=-0.5,~~~\beta_{\omega p}=0.1,\nonumber\\
	\Delta&=&\frac{\pi}{3},~~~~\Delta_{01}=\frac{\pi}{4},~~~~~~\Delta_{10}=\frac{\pi}{6},
\end{eqnarray}
as well as considering different $\beta_{\omega p}$, $\gamma_{\omega p}$ values, the sensitivities versus the statistics are shown in Fig.~\ref{fig:beta} and  Fig.~\ref{fig:gamma}.

\section{summary and outlook}
To study the weak interactions in the $\ldc\to\phi p$ and $\ldc\to\omega p$ decays, we formulate the asymmetry parameters in these two decays by using the helicity amplitudes, and the joint angular distributions are provided. Specially to investigate the asymmetry parameters at the BESIII and the LHCb experiments, we show how the polarization is transferred from the mother particles to the daughter particles. By making use of the full decay chain, the sensitivity of measuring the asymmetry parameters is estimated for the channel $\ldc\to\omega p, \omega\to\pi^0\pi^+\pi^-$ as well as $\ldc\to\phi p, \phi \to K^+ K^-$. The estimation is very helpful to the physics program design in the near future facilities, such as CEPC, STCF etc.

\section{acknowledgments}
The work is partly supported by the National Natural Science Foundation of China under Grants No. 12175244, No. 11875262, No. 11805037 and No. U1832121. National Key Research and Development Program of China under Contracts No. 2020YFA0406301.
\appendix
\section{THE ELEMENTS OF  $\phi$ SDM}\label{SDM of phi}
\begin{eqnarray}
	\rho^{\phi}_{1,1}&=&\frac{1}{2} \bm{\mathcal{P}_0}  (1+\bm{\mathcal{P}_y} \mathrm{\sin}\theta _1\mathrm{\sin} \phi_1)b_{1,\frac{1}{2}}^2,\nonumber\\
	\rho^{\phi}_{1,0}&=&\frac{1}{2} \bm{\mathcal{P}_0} \bm{\mathcal{P}_y}  e^{-i \Delta_{01}} (\mathrm{\cos} \theta_1 \mathrm{\sin} \phi_1-i \mathrm{\cos}\phi_1)b_{0,\frac{1}{2}} b_{1,\frac{1}{2}},\nonumber\\
	\rho^{\phi}_{1,-1}&=&0,\nonumber\\
	\rho^{\phi}_{0,1}&=&\frac{1}{2} \bm{\mathcal{P}_0} \bm{\mathcal{P}_y} e^{i \Delta_{01}} (\mathrm{\cos} \theta _1 \mathrm{\sin} \phi_1+i \mathrm{\cos}\phi_1)b_{0,\frac{1}{2}} b_{1,\frac{1}{2}} ,\nonumber\\
	\rho^{\phi}_{0,0}&=&\frac{1}{2} \bm{\mathcal{P}_0}[(1+\bm{\mathcal{P}_y}\mathrm{\sin}\theta_1\mathrm{\sin}\phi_1)b_{0,\frac{1}{2}}^2\nonumber\\
								 &+&(1-\bm{\mathcal{P}_y} \mathrm{\sin}\theta_1 \mathrm{\sin}\phi_1)b_{0,-\frac{1}{2}}^2 ],\nonumber\\
	\rho^{\phi}_{0,-1}&=&\frac{1}{2} \bm{\mathcal{P}_0} \bm{\mathcal{P}_y} e^{-i \Delta_{10}} (\mathrm{\cos}\theta_1 \mathrm{\sin}\phi_1-i \mathrm{\cos}\phi_1)b_{-1,-\frac{1}{2}} b_{0,-\frac{1}{2}},\nonumber\\
	\rho^{\phi}_{-1,1}&=&0,\nonumber\\
	\rho^{\phi}_{1,1}&=&\frac{1}{2} \bm{\mathcal{P}_0} \bm{\mathcal{P}_y}  e^{i \Delta_{10}} (\mathrm{\cos}\theta_1 \mathrm{\sin}\phi_1+i \mathrm{\cos} \phi_1)b_{-1,-\frac{1}{2}} b_{0,-\frac{1}{2}},\nonumber\\
	\rho^{\phi}_{1,1}&=&\frac{1}{2} \bm{\mathcal{P}_0}(1-\bm{\mathcal{P}_y}\mathrm{\sin}\theta_1 \mathrm{\sin}\phi_1)b_{-1,-\frac{1}{2}}^2,\nonumber
\end{eqnarray}

\section{REAL MULTIPOLE PARAMETERS OF $\ldc$ IN $\ldc \to \phi p$ DECAY}\label{mul_para}
\begin{eqnarray*}
	2r^0_0&=&{\bm{\mathcal{P}_0}}\left\{\right.b_{-1,-\frac{1}{2}}^2 (1-\bm{\mathcal{P}_z}\mathrm{\cos}\theta_1-\bm{\mathcal{P}_x}\mathrm{\cos}\phi_1\mathrm{\sin}\theta_1\nonumber\\
				 &-&\bm{\mathcal{P}_y}\mathrm{\sin}\theta_1\mathrm{\sin}\phi_1)+b_{0,-\frac{1}{2}}^2+b_{0,\frac{1}{2}}^2+b_{1,\frac{1}{2}}^2\nonumber\\
				 &+&[\bm{\mathcal{P}_z}\mathrm{\cos}\theta_1+\mathrm{\sin}\theta_1(\bm{\mathcal{P}_x}\mathrm{\cos}\phi_1+\bm{\mathcal{P}_y}\mathrm{\sin}\phi_1)]\nonumber\\
				 &\times&(b_{0,-\frac{1}{2}}^2-b_{0,\frac{1}{2}}^2+b_{1,\frac{1}{2}}^2)\left.\right\},\nonumber\\
		\end{eqnarray*}

\begin{eqnarray}
	r^0_0r^1_{-1}&=&\frac{1}{2}\sqrt{3\over2}{\bm{\mathcal{P}_0}}\left\{\right. 		[\mathrm{\cos}\Delta_{10}(\bm{\mathcal{P}_y}\mathrm{\cos}\phi_1-\bm{\mathcal{P}_x}\mathrm{\sin}\phi_1)\nonumber\\
	&+&(-\bm{\mathcal{P}_z}\mathrm{\sin}\theta_1
							+\mathrm{\cos}\theta_1(\bm{\mathcal{P}_x}\mathrm{\cos}\phi_1+\bm{\mathcal{P}_y}\mathrm{\sin}\phi_1))\nonumber\\
							&\times& \mathrm{\sin}\Delta_{10} ]b_{-1,-\frac{1}{2}}b_{0,-\frac{1}{2}} +[\mathrm{\cos}\Delta_{01}(\bm{\mathcal{P}_y}\mathrm{\cos}\phi_1\nonumber\\
							&-&\bm{\mathcal{P}_x}\mathrm{\sin}\phi_1)+(-\bm{\mathcal{P}_z}\mathrm{\sin}\theta_1+\mathrm{\cos}\theta_1(\bm{\mathcal{P}_x}\mathrm{\cos}\phi_1\nonumber\\
							&+&\bm{\mathcal{P}_y}\mathrm{\sin}\phi_1))\mathrm{\sin}\Delta_{01}]b_{0,\frac{1}{2}}b_{1,\frac{1}{2}}\left.\right\}\nonumber\\
	r^0_0r^1_{0}&=&\frac{\sqrt{3}}{4}{\bm{\mathcal{P}_0}}[(-1+\bm{\mathcal{P}_z}\mathrm{\cos}\theta_1+\bm{\mathcal{P}_x}\mathrm{\cos}\phi_1\mathrm{\sin}\theta_1\nonumber\\
						   &+&\bm{\mathcal{P}_y}\mathrm{\sin}\theta_1\mathrm{\sin}\phi_1)b_{-1,-\frac{1}{2}}^2+(1+\bm{\mathcal{P}_z}\mathrm{\cos}\theta_1\nonumber\\
						   &+&\bm{\mathcal{P}_x}\mathrm{\cos}\phi_1\mathrm{\sin}\theta_1+\bm{\mathcal{P}_y}\mathrm{\sin}\theta_1\mathrm{\sin}\phi_1)b_{1,\frac{1}{2}}^2]\nonumber\\
	r^0_0r^1_{1}&=&\frac{1}{2}\sqrt{3\over2}{\bm{\mathcal{P}_0}}\left\{\right. [\mathrm{\cos}\Delta_{10}(-\bm{\mathcal{P}_z}\mathrm{\sin}\theta_1\nonumber\\
							&+&\mathrm{\cos}\theta_1(\bm{\mathcal{P}_x}\mathrm{\cos}\phi_1+\bm{\mathcal{P}_y}\mathrm{\sin}\phi_1))\nonumber\\
							&-&(\bm{\mathcal{P}_y}\mathrm{\cos}\phi_1-\bm{\mathcal{P}_x}\mathrm{\sin}\phi_1)\mathrm{\sin}\Delta_{10} ]b_{-1,-\frac{1}{2}}b_{0,-\frac{1}{2}}\nonumber\\ &+&[\mathrm{\cos}\Delta_{01}(-\bm{\mathcal{P}_z}\mathrm{\sin}\theta_1+\mathrm{\cos}\theta_1(\bm{\mathcal{P}_x}\mathrm{\cos}\phi_1\nonumber\\
							&+&\bm{\mathcal{P}_y}\mathrm{\sin}\phi_1))-(\bm{\mathcal{P}_y}\mathrm{\cos}\phi_1-\bm{\mathcal{P}_x}\mathrm{\sin}\phi_1)\mathrm{\sin}\Delta_{01}]\nonumber\\
							&\times&b_{0,\frac{1}{2}}b_{1,\frac{1}{2}}\left.\right\}\nonumber\\
	r^0_0r^2_{-1}&=&\frac{1}{2}\sqrt{3\over2}{\bm{\mathcal{P}_0}}\left\{\right.-[\mathrm{\cos}\Delta_{10}(\bm{\mathcal{P}_y}\mathrm{\cos}\phi_1-\bm{\mathcal{P}_x}\mathrm{\sin}\phi_1)\nonumber\\
							 &+&(-\bm{\mathcal{P}_z}\mathrm{\sin}\theta_1+\mathrm{\cos}\theta_1(\bm{\mathcal{P}_x}\mathrm{\cos}\phi_1+\bm{\mathcal{P}_y}\mathrm{\sin}\phi_1))\nonumber\\
							 &\times& \mathrm{\sin}\Delta_{10} ]b_{-1,-\frac{1}{2}}b_{0,-\frac{1}{2}} +[\mathrm{\cos}\Delta_{01}(\bm{\mathcal{P}_y}\mathrm{\cos}\phi_1\nonumber\\
							 &-&\bm{\mathcal{P}_x}\mathrm{\sin}\phi_1)
							 +(-\bm{\mathcal{P}_z}\mathrm{\sin}\theta_1+\mathrm{\cos}\theta_1(\bm{\mathcal{P}_x}\mathrm{\cos}\phi_1\nonumber\\
							 &+&\bm{\mathcal{P}_y}\mathrm{\sin}\phi_1))
							 \times \mathrm{\sin}\Delta_{01}]b_{0,\frac{1}{2}}b_{1,\frac{1}{2}}\left.\right\}\nonumber\\
	r^0_0r^2_0&=&\frac{1}{4}{\bm{\mathcal{P}_0}}[-(-1+\bm{\mathcal{P}_z}\mathrm{\cos}\theta_1+\bm{\mathcal{P}_x}\mathrm{\cos}\phi_1\mathrm{\sin}\theta_1\nonumber\\
						 &+&\bm{\mathcal{P}_y}\mathrm{\sin}\theta_1\mathrm{\sin}\phi_1)b_{-1,-\frac{1}{2}}^2-2(b_{0,-\frac{1}{2}}^2+b_{0,\frac{1}{2}}^2)\nonumber\\
						 &+&b_{1,\frac{1}{2}}^2
						 -(\bm{\mathcal{P}_z} \mathrm{\cos}\theta_1+\mathrm{\sin}\theta_1(\bm{\mathcal{P}_x}\mathrm{\cos}\phi_1\nonumber\\
						 &+&\bm{\mathcal{P}_y}\mathrm{\sin}\phi_1))
						 \times(2b_{0,-\frac{1}{2}}^2-2b_{0,\frac{1}{2}}^2-b_{1,\frac{1}{2}}^2)
	]\nonumber\\
	r^0_0r^2_1&=& \frac{1}{2}\sqrt{3\over2}{\bm{\mathcal{P}_0}}\left\{\right. -[\mathrm{\cos}\Delta_{10}(-\bm{\mathcal{P}_z}\mathrm{\sin}\theta_1\nonumber\\
	&+&\mathrm{\cos}\theta_1(\bm{\mathcal{P}_x}\mathrm{\cos}\phi_1
						 +\bm{\mathcal{P}_y}\mathrm{\sin}\phi_1))-(\bm{\mathcal{P}_y}\mathrm{\cos}\phi_1\nonumber\\
						 &-&\bm{\mathcal{P}_x}\mathrm{\sin}\phi_1)\mathrm{\sin}\Delta_{10} ]b_{-1,-\frac{1}{2}}b_{0,-\frac{1}{2}}\nonumber\\ &+&[\mathrm{\cos}\Delta_{01}(-\bm{\mathcal{P}_z}\mathrm{\sin}\theta_1+\mathrm{\cos}\theta_1(\bm{\mathcal{P}_x}\mathrm{\cos}\phi_1\nonumber\\
						 &+&\bm{\mathcal{P}_y}\mathrm{\sin}\phi_1))
						-(\bm{\mathcal{P}_y}\mathrm{\cos}\phi_1-\bm{\mathcal{P}_x}\mathrm{\sin}\phi_1)\mathrm{\sin}\Delta_{01}]\nonumber\\
						&\times&b_{0,\frac{1}{2}}b_{1,\frac{1}{2}}\left.\right\}\nonumber\\
	r^0_0r^2_2&=&	r^0_0r^2_{-2}=0.\nonumber
\end{eqnarray}


\end{document}